\documentclass[aps,prl,twocolumn,nopacs,floats,superscriptaddress]{revtex4-1}
\usepackage{graphicx}
\usepackage{dcolumn}
\usepackage{bm}
\usepackage{amsmath}
\usepackage{color}

\definecolor{cardinal}{rgb}{0.6,0,0}
\definecolor{darkgreen}{rgb}{0,0.4,0}
\definecolor{golden}{rgb}{0.92, 0.7, 0}
\definecolor{midnight}{rgb}{0, 0, 0.5}
\definecolor{darkblue}{rgb}{0, 0, 0.7}

\def\he4{$^4$He}
\def\hel3{$^3$He}
\def\Am3{\AA$^{-3}$}
\def\beq{\begin{equation}}
\def\eeq{\end{equation}}

\newcommand{\pt}{{\partial}}

\newcommand{\rv}{{\bf r}}

\newcommand{\Jh}{{\hat{J}}}
\newcommand{\ah}{{\hat a}}
\newcommand{\nh}{{\hat n}}

\newcommand{\oh}{{\frac{1}{2}}}

\newcommand{\as}{\bar a}

\newcommand{\be}{\begin{equation}}
\newcommand{\ee}{\end{equation}}
\newcommand{\bea}{\begin{eqnarray}}
\newcommand{\eea}{\end{eqnarray}}
\newcommand{\bse}{\begin{subequations}}
\newcommand{\ese}{\end{subequations}}
\def\rf#1{(\ref{#1})}
\def\rfs#1{Eq.~\rf{#1}}

\begin{document}

\author{Leo Radzihovsky}
\email{radzihov@colorado.edu}

\author{Emil Pellett}
\affiliation{Department of Physics and Center for Theory of Quantum
  Matter, University of Colorado, Boulder, CO 80309, USA}


\title{Quantum phases and  transitions of bosons on a comb lattice}

\begin{abstract}
  Motivated to elucidate the nature of quantum phases and their
  criticality when entangled with a correlated quantum bath, we study
  interacting bosons on a ``comb lattice'' -- a one-dimensional
  backbone (system) coupled at its sites to otherwise independent
  one-dimensional ``teeth'' chains (bath). We map out the
  corresponding phase diagram, detailing the nature of the phases and
  phase transitions. Controlled by the backbone and teeth hopping
  amplitudes, short-range interactions and chemical potential, phases
  include a Mott-insulator (MI), backbone (LL$_b$) and teeth
  (LL$_\perp$) Luttinger liquids, and the long-range ordered
  incoherent superfluid (iSF). We explore their properties and
  potential realizations in condensed matter and cold-atom experiments
  and simulations.
\end{abstract}

\maketitle

{\em Introduction and motivation.}---Behavior of a quantum system
coupled to an environment remains an important problem in physics. It
informs us on fundamental and applied questions of decoherence,
measurement and the emergence of the deterministic classical world
from the ephemeral and uncertain quantum one.\cite{CLprl81, CLaop83,
  CLphysicaA83, DecoherenceClassicaQuantumlWorldBook} Recently, these
questions have become particularly vexing, in the context of quantum
simulation and computing, performed in a large variety of
well-engineered, yet noisy (NISQ) quantum systems, where they can be
experimentally explored in great
detail.\cite{BrandtQC1999,UnruhQCpra1995,DiVincenzoScience1995} Other
realizations arise in surface or low-dimensional subsystems, coupled
to their nontrivial
bulk~\cite{Burovskii,dislocationPRL,iTQFprb,fingerprintsPRA},
including critical~\cite{MaxMetlitskiPRL22, RevisitedMaxMetlitski21}
and Symmetry Protected Topological (SPT)
phases~\cite{GroverVishwanathSPIT}, as well as a variety of
polaron-like problems~\cite{FeynmanPolaron, ZimanyiPRB87,
  NetoFisher,KunYangEdge,NayakOpenLL,Pankov,Werner,Cai,LodeQMC,
  Weber,Danu,Martin,LRunpublished}, where gapless Goldstone (e.g.,
phonons, Bogoliubov modes, and spin-waves), critical or Fermi-surface
modes play the role of a bath.

In most studies addressing such problems, dating back to the seminal
work of Caldeira and Leggett, the environment is phenomenologically
modeled by a low-energy spectrum of quantum harmonic
oscillators~\cite{CLprl81,CLphysicaA83,CLaop83,ChakravartyLeggettPRL84,SchmidPRL83,GuineaPRL85,WeissPhysLett85,FisherZwergerPRB85,LeggettRMP,ZimanyiPRB87,
  ZimanyiPRB88, ZimanyiPRL90, ZimanyiPRB93, Weber, WeberPRB24,
  RossoPRB2024}. Its only role is to generate decoherence and
dissipation for the quantum system coupled to it, but featureless and
uninteresting otherwise.  In constrast, however, many physical
systems, such as e.g., Josephson junction arrays and bosons in optical
lattices can be far richer, with both the system and the environment
strongly correlated, and exhibiting a variety of phases and associated
quantum phase transitions.

\begin{figure}[!htb]
\hspace*{-0.3cm}\includegraphics[width=1.1
\columnwidth]{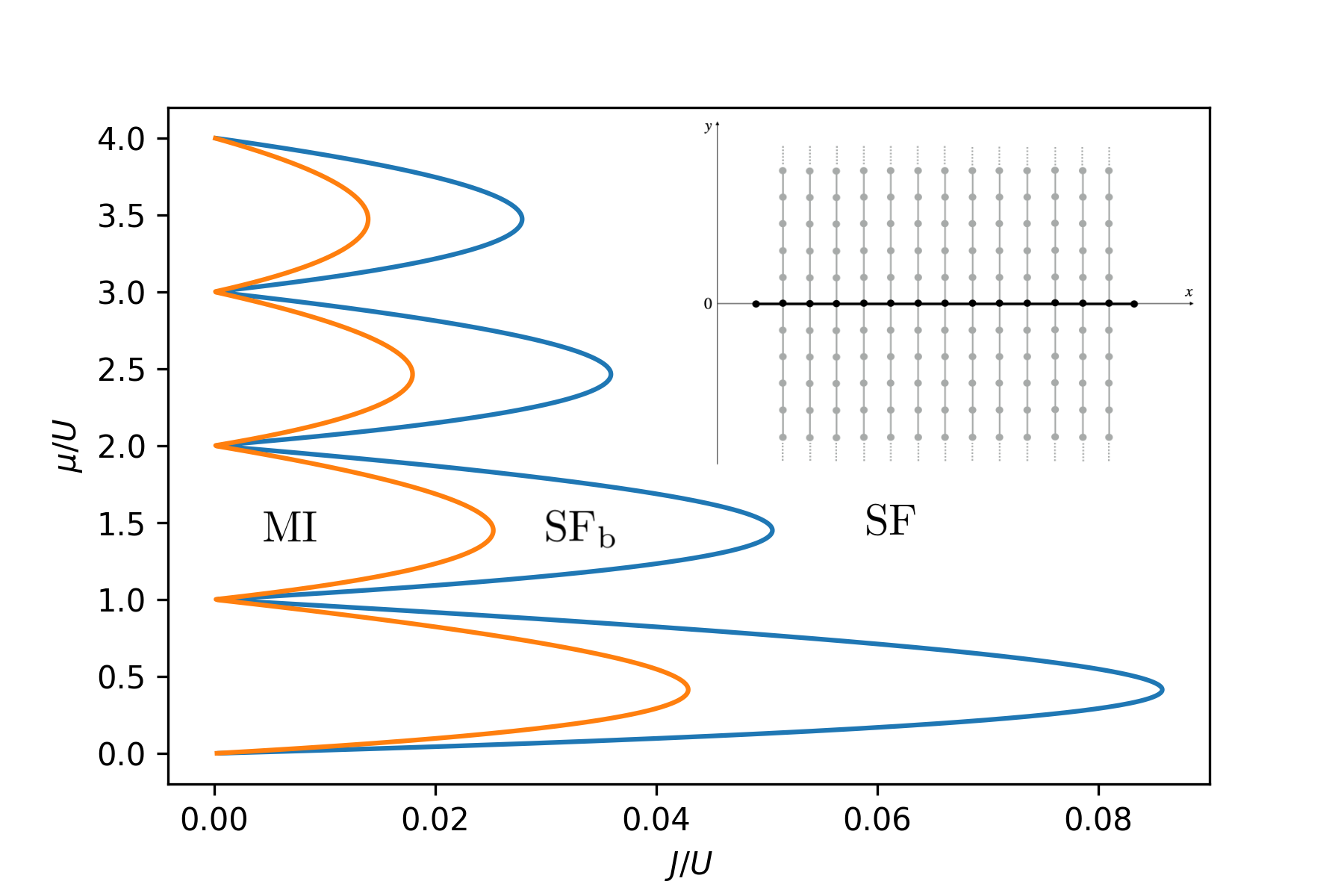}
\caption{(Inset) A ``fish-bone'' (equivalently ``comb'') lattice,
  where vertical ``bath'' bosonic chains are coupled to the horizontal
  ``system'' chain at $y=0$, but are otherwise decoupled (adapted from
  Ref.~\onlinecite{iTQFprb}). The backbone chain (in black) can also
  be viewed as a coupled chain of Kane-Fisher dots \cite{KaneFisher}.
  (Main) Chemical potential ($\mu$) versus hopping amplitude
  ($J = J_\perp$, backbone and teeth, respectively) mean-field (MF)
  phase diagram, illustrating incompressible Mott-Insulator,
  backbone-only superfluid SF$_b$ and bulk and backbone are in a
  superfluid SF phase (to our knowledge first studied via MF theory
  (MFT) and QMC in Ref.\onlinecite{BuonsanteMFT_QMC_PRB2004}).}
\label{latticeFig}
\end{figure}

To explore such systems, here we study a Bose-Hubbard (BH) model with
bosons restricted to hop on a ``comb'' (or, equivalently, a
``fish-bone'' \cite{LodeQMC,iTQFprb, fingerprintsPRA,
  BuonsanteMFT_QMC_PRB2004}) lattice (see Fig.\ref{latticeFig}), that
can be efficiently simulated with Quantum Monte Carlo
(QMC)\cite{ZimanyiPRL90, LodeQMC, fingerprintsPRA,WeberPRB24,
  BuonsanteMFT_QMC_PRB2004}, and, for spins, with Density Matrix
Renormalization Group (DMRG) methods~\cite{WhiteQMC}.  In such
geometry it is natural to view the one-dimensional (1D) backbone
(x-chain) of the comb lattice as the system (S) and the array of
transverse 1d comb teeth ($\perp$-chains) as the environment or a
bath.

For hopping amplitudes large compared to the repulsive interaction
(and/or incommensurate lattice filling), this system has been
extensively explored in Refs.~\onlinecite{LodeQMC, iTQFprb,
  fingerprintsPRA}, demonstrating that at zero temperature it exhibits
long-range (LR) incoherent superfluid (iSF) order (dubbed ``incoherent
Transverse Quantum Fluid'' (iTQF) in Ref.~\onlinecite{iTQFprb}),
despite the one-dimensionality of the backbone and otherwise decoupled
1d teeth.  The iSF phase has also been demonstrated in QMC simulations
of a model of 1d bosons coupled to an ohmic bath of harmonic
Caldeira-Leggett oscillators.\cite{WeberPRB24} In this novel phase,
the quasi-one-dimensional LR SF order is enabled by the
quasi-long-range (QLR) SF order of the teeth (transverse chains),
contrasting with localizing effects of the bath in
Ref.~\onlinecite{FisherZwergerPRB85}.  As discussed in
Ref. \onlinecite{iTQFprb}, the iSF exhibits a divergent
compressibility with enhanced boson number fluctuations into the
transverse Luttinger liquid (LL) chains, which suppresses SF phase
fluctuations.  One can see this by integrating out bath-bosons on
these $\perp$-chains, reducing to an effective ``Kane-Fisher''
backbone x-chain \cite{KaneFisher}, whose linear-in-$\omega$ Gaussian
action kernel, $\Gamma_{\omega,k}=|\omega| + n_sk^2$, leads to finite
superfluid phase fluctuations.  The associated long-range superfluid
order of the one-dimensional iSF (iTQF) contrasts qualitatively with
isolated incommensurate 1d bosons (i.e., in the absence of
$\perp$-chains), forming a quasi-long-range SF ordered Luttinger
liquid~\cite{GiamarchiBosonizationBook}.

Although this novel iSF state is quite well understood~\cite{iTQFprb,
  fingerprintsPRA}, it is natural to ask about the full phase diagram
of bosons on such comb lattice, and to explore corresponding phases
that iSF can transition to and the nature of the associated critical
phenomena, some of which has been investigated in a QMC
study~\cite{LodeQMC}.

%
%

\begin{figure}[!htb]
  \includegraphics[width=0.8 \columnwidth]{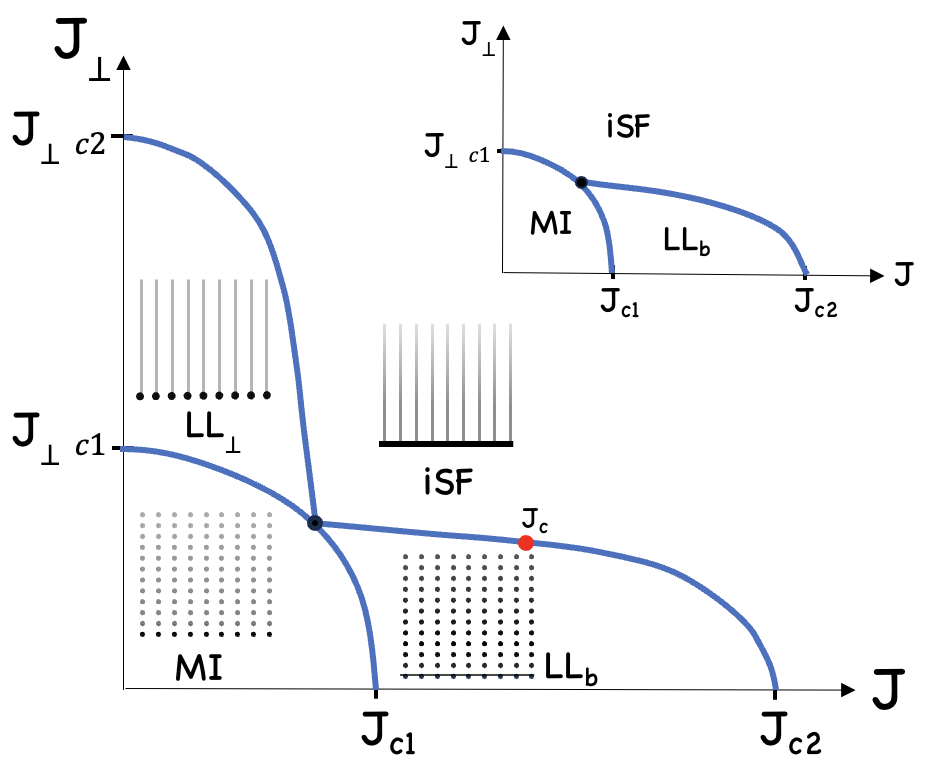}
  \caption{Schematic phase diagram for hopping amplitudes
    $\Jh_\perp - \Jh$ at fixed chemical potential $\hat\mu$ (in units
    of repulsive interaction $U$, generically allowing beyond on-site
    interaction) for bosons on a comb (or, equivalent fish-bone)
    lattice. For low-commensurability $\nu = q/p$, $p > 2$ lattice
    filling, the system displays (i) gapped Mott-Insulator (MI), (ii)
    a backbone Luttinger liquid (LL$_b$), with QLRO along the
    backbone, and short-range correlated and insulating transverse
    teeth $\perp$-chains, (iii) transverse Luttinger liquid
    (LL$_\perp$), with QLRO along decoupled transverse teeth
    $\perp$-chains and short-ranged along the backbone, and (iv)
    incoherent Superfluid (iSF), characterized by LR SF order along
    the 1D backbone at $y = 0$ and QLR SF order along the transverse
    LL teeth. The dot (red online) at $J_c$ on the LL$_b$-iSF boundary
    separates it into extraordinary-LL and extraordinary-log
    transitions~\cite{MaxMetlitskiPRL22, RevisitedMaxMetlitski21}.  The
    images schematically indicate superfluid order parameter
    correlations. Inset displays the phase diagram for
    high-commensurability filling for $p \le 2$, where the LL$_\perp$
    phase merges with the iSF.}
\label{phasediagramFig}
\end{figure}

{\em Results.}---Before turning to a detailed analysis, we summarize
our key results. Treating the Bose-Hubbard model on a comb lattice
using mean-field, pertubative and renormalization-group analyses we
derive universal features of its zero-temperature phase diagram as a
function of the normalized chemical potential $\hat\mu\equiv\mu/U$ and
transverse, $\hat J_\perp\equiv J_\perp/U$ and backbone,
$\hat J \equiv J/U$ hopping amplitudes, in units of on-site repulsive
Hubbard $U$ interaction.

A first principles analysis (e.g., via QMC) is needed for the
prediction of the quantitative details of the global phase diagram,
that we hope our work
stimulates\cite{ChaoZhangQMCunpublished}. However, qualitative
structure and universal features (robust to weak perturbations to our
model), can be deduced on general grounds and RG analysis. As
illustrated in Fig.\ref{phasediagramFig}, for beyond on-site
interaction, and backbone and teeth rational fillings, $\nu = q/p$ and
$\nu_\perp = q_\perp/p_\perp$, respectively~\cite{commentFilling} with
$p > 2$, $p_\perp > 2$, it generically displays four distinct phases:
(i) gapped Mott insulator (MI) for commensurate lattice filling and
small hopping
$\Jh < \Jh_{c1} \sim O(1) ,\ \ \Jh_\perp < \Jh_{\perp,c1} \sim O(1)$,
(ii) gapless ``backbone Luttinger liquid'' (LL$_b$), characterized by
quasi-long-range order (QLRO) along the backbone and short-range
(exponential decay) off-diagonal order transverse to the backbone for
$\Jh_{c1} < \Jh < \Jh_{c2},\ \ \Jh_\perp < \Jh_{\perp,c1}$, (iii)
gapless ``transverse Luttinger liquid'' (LL$_\perp$), characterized by
quasi-long-range order (QLRO) along decoupled transverse chains (comb
teeth) and short-range off-diagonal order along the backbone for
$\Jh < \Jh_{c1},\ \ \Jh_{\perp,c1}<\Jh_\perp < \Jh_{\perp,c2}$, (iv)
gapless iSF, exhibiting ODLR superfluid order along the (system)
backbone chain and QLRO (bath) teeth chains transverse to it for
$\Jh > \Jh_{c2},\ \ \Jh_\perp > \Jh_{\perp,c2}$.

For $p_\alpha = 2$ ($\nu_\alpha = 1/2$) and integer filling
$\nu_\alpha = n$ (an inset in Fig.~\ref{phasediagramFig}) a distinct
LL$_\perp$ phase is not allowed, and is instead smoothly connecting to
iSF, as QLR SF order of the transverse chains (bulk) necessarily
induces a relevant interchain hopping, leading to LR SF order along
the backbone.

The MI-LL$_{b}$ and MI-LL$_\perp$ quantum phase transitions at
$\Jh_{c1}$ and $\Jh_{\perp, c1}$, respectively are of the standard
Kosterlitz-Thouless (KT) universality class. The system also allows a
direct continuous MI-iSF transition for $\Jh \approx \Jh_{\perp}$
through a tetracritical point.  Alternatively, the two LL phases can
be separated by a first-order LL$_b$-LL$_\perp$ transition, or the
tetracritical point can be replaced by a first-order MI-iSF boundary,
scenarios that we cannot exclude, requiring numerical, e.g., QMC
treatment.\cite{ChaoZhangQMCunpublished} The LL$_{b}$-iSF quantum
phase transition at $\Jh_{c2}(\Jh_\perp)$ is of an unusual {\em
  extra-ordinary} universality class~\cite{MaxMetlitskiPRL22,
  RevisitedMaxMetlitski21}, where the bulk (transverse teeth chains)
transitions from SRO to QLRO in the presence of the QLRO boundary
(backbone chain), which subsequently develops LRO inside the iSF. We
leave the analysis of this novel quantum criticality to a future
study. Finally, we predict that the LL$_\perp$-iSF quantum phase
transition at $\Jh_{\perp,c2}$ is of a Kosterlitz-Thouless type, where
hopping along the backbone at $y=0$, in the presense of the transverse
QLRO, becomes relevant above $\Jh_{\perp,c2}(\Jh)$. We characterize
these phases by their correlation along the back-bone and transverse
to it (along the comb teeth). For irrational fillings, only the iSF
phase is stable.

{\em Model.}--- We utilize the standard Bose-Hubbard model for lattice
bosons created by $\ah^\dagger_{x,y}$ at site $\rv = (x,y)$, with the
grand-canonical Hamiltonian,
\bea
\hat H &=&-J\sum_x(\ah^\dagger_{x,0} \ah_{x+1,0} + h.c.)
- J_\perp\sum_{x,y} (\ah^\dagger_{x,y} \ah_{x,y+1}  + h.c.)\nonumber\\
&&-\mu\sum_{\rv}\nh_{\rv} + \oh U\sum_{\rv}\nh_{\rv}(\nh_{\rv}-1).
\label{H}
\eea
Boson on-site number $\nh_\rv = \ah^\dagger_{\rv} \ah_{\rv}$ is (on
average) controlled by the chemical potential $\mu$, $U$ is on-site
repulsion (for simplicity taken to be the same on all
sites)~\cite{commentOnSiteU} and $J, J_\perp$ are, respectively, the
hopping matrix elements along the backbone x-chain at $y=0$ and along
the transverse $\perp$-chains (comb teeth), as illustrated in
Fig.\ref{latticeFig}.  Aside from the commensurate filling factors,
$\nu_\alpha$, our predictions are based on universal field-theoretic
approach and are thus quite generic and robust to weak perturbations
to the model, \rfs{H}. The partition function
$Z = \text{Tr}\big [e^{-\beta \hat H}\big ] = \int[da d \as d\psi
d\bar\psi d\Psi d \bar\Psi]e^{-\int_0^\beta d\tau L}$ is conveniently
expressed as a coherent-state ($\hat a|a\rangle = a|a\rangle$)
path-integral, with the Euclidean Lagrangian,
\bea L &=& \sum_\rv\left[\as_\rv\pt_\tau a_\rv -\mu \as_\rv a_\rv
  +\frac{U}{2} \as_\rv \as_\rv a_\rv a_\rv\right]\nonumber\\
&&+ \sum_x(\psi_x \as_{x,0} +   c.c.)
+ \sum_\rv(\Psi_\rv \as_\rv +   c.c.)
\nonumber\\
&&+ \sum_{x,x'}\bar\psi_x J^{-1}_{x,x'}\psi_{x'}
+ \sum_{\rv,\rv'}\bar\Psi_\rv J^{-1}_{\perp,\rv,\rv'}\Psi_{\rv'}\;.
\label{L1}
\eea
Above we introduced two Hubbard-Stratonovic fields, $\psi_x$ and
$\Psi_\rv$ to decouple boson hopping along $x$ on the backbone chain at
$y=0$ and along transverse $y$-chains at each $x$, with $J^{-1}_{x,x'}$ and
$J^{-1}_{\perp,\rv,\rv'}=J^{-1}_{\perp,y,y'}\delta_{x,x'}$ the
inverses of the corresponding hopping matrix elements, respectively.

Perturbatively integrating out the microscopic degrees of freedom,
$a_\rv(\tau)$, gives $Z =\int[d\psi d\bar \psi d\Psi d\bar
\Psi]e^{-S}$, with the effective action,
\begin{widetext}
\bea 
S &=&
\int_{\tau,\tau'}\bigg[
\sum_{\rv,\rv'}\bar\Psi_{\rv,\tau}
  \Gamma_{\perp,\rv,\rv'}(\tau-\tau')\Psi_{\rv',\tau'} +\sum_{x,x'}
\left[\bar\psi_{x,\tau}\Gamma_{x,x'}(\tau-\tau')\psi_{x',\tau'} 
  - \bar\psi_{x,\tau} C(\tau-\tau')\Psi_{x,0,\tau'}+c.c.\right]
+\ldots \bigg].
\eea
\end{widetext}
%
%
where $\ldots$ represent quartic and higher order terms in $\Psi$ and
$\psi$.  The corresponding kernels are given by,
\bea
\Gamma_{\perp,\rv,\rv'}(\tau-\tau')&=&
J^{-1}_{\perp,y, y'}\delta_{x,x'}\delta(\tau-\tau') - \delta_{\rv,\rv'}C(\tau-\tau'),\;\;\;\;\;\;\\
\Gamma_{x,x'}(\tau-\tau')&=&
J^{-1}_{x,x'}\delta(\tau-\tau') -  \delta_{x,x'}C(\tau-\tau'),\\
C(\tau-\tau')&=& \langle a_\rv(\tau) \bar a_{\rv}(\tau')\rangle_0,
\eea
where the averages are uncorrelated in $\rv$, taken with the local
part of the action in the first line of $L$ in \rf{L1}, i.e., with
$\Psi = \psi = 0$.

Expanding at low frequencies and momenta, the leading part of the
low-energy action is given by
\begin{widetext}
\bea 
S &=&
\int_{\tau,\rv}\bigg[\gamma_{1\perp}\bar\Psi\partial_\tau\Psi
+\gamma_{2\perp}|\partial_\tau\Psi|^2
+ K_\perp|\partial_y\Psi|^2 + a_\perp|\Psi|^2
+\oh b_\perp|\Psi|^4\nonumber\\
&&+\delta(y)\big(\gamma_1\bar\psi\partial_\tau\psi
+\gamma_2|\partial_\tau\psi|^2
+ K|\partial_x\psi|^2 + a|\psi|^2+\oh b|\psi|^4
-c \bar\psi\Psi + c.c. +\ldots\big)\bigg],\nonumber\\
\label{Sfull}
\eea
\end{widetext}
where above couplings can be directly derived as zero frequency, local
on-site correlators of the microscopic model \rf{L1} for
$\Psi = \psi = 0$.  Importantly, the critical couplings
$a_\alpha = (a_\perp, a)$
are given by
\bea
a_\alpha&=&\frac{1}{z J_\alpha} - \frac{\hat\chi_0(\mu/U)}{U},
\label{a_alpha}
\eea
where,
\bea
\hat\chi_0(\hat\mu) &=&c  = \tilde C(\omega = 0),\\
&=&\frac{n_0(\hat\mu) +
    1}{n_0(\hat\mu) - \hat\mu} + \frac{n_0(\hat\mu)}{\hat\mu +1 -
    n_0(\hat\mu)}\\
&=&\frac{\hat\mu+1}{(\hat\mu_I+1 -\hat\mu) (\hat\mu - \hat\mu_I)},
\eea
$n_0(\hat\mu) = \hat\mu_I + 1$ is the boson number per site and
$\hat\mu_I$ is the integer part of
$\hat\mu \equiv \mu/U$~\cite{SachdevBook,Gurarie}. The precise form of
all other couplings is unimportant to us, except that they are
positive, with $\gamma_\alpha = -\partial a_\alpha/\partial\mu$,
vanishing at the particle-hole symmetric point of the Mott insulator
lobes, defined below.\cite{SachdevBook}

{\em High-dimensional model.}---To get a sense of the phase diagram,
it is useful to first consider a higher $d$ dimensional generalization
of the 2+1D comb lattice model \rf{Sfull}, which allows long-range
U(1) symmetry-broken orders (with transverse teeth and backbone each
higher than one- (two-) dimensional at $T=0$ ($T>0$)) and validity of
MFT treatment\cite{BuonsanteMFT_QMC_PRB2004}. With this generalization
and (at first) in the absence of the coupling $c$, i.e., $c=0$, there
are $4$ distinct phases, corresponding to: (i) Mott Insulator (MI) for
$a > 0, a_\perp > 0$, with $\psi = 0, \Psi = 0$, (ii) Backbone
Superfluid (SF$_{b}$) for $a < 0, a_\perp > 0$, with
$\psi \neq 0, \Psi = 0$, (iii) Transverse Superfluid (SF$_\perp$) for
$a > 0, a_\perp < 0$, with $\psi = 0, \Psi \neq 0$, (iv) Superfluid
(SF) $a < 0, a_\perp < 0$, with $\psi\neq 0, \Psi \neq 0$.  However,
in these higher dimensions $c\neq 0$ is always relevant and thereby
hybridizes $\Psi$ and $\psi$, corresponding to hopping of bosons
between backbone and transverse teeth at $y=0$.  This enforces a
nonzero $\psi$ for any $\Psi \neq 0$, corresponding to the observation
that a surface ($y=0$ backbone x-chain) of a bulk superfluid
(transverse teeth) is also necessarily a superfluid. Thus, in this
high-dimensional generalization, this nonzero $c$ generically
eliminates above case (iii) $\psi = 0, \Psi \neq 0$, SF$_\perp$ as a
separate phase.  In MFT the MI - SF$_{b}$ lobe boundary is now defined
by $a = c^2/a_\perp$, suppressing the MI by the coupling $c$ to the
bulk (comb teeth).  The resulting $3$ phases are separated by
$d_\perp$ and $d-d_\perp$ dimensional MI-SF quantum phase transitions
(with two distinct types, depending on whether $\gamma_\alpha = 0$
[particle-hole symmetric] or not) as $J_\alpha/U$ are tuned for fixed
$\mu$~\cite{SachdevBook}. For $J_\perp/J \gg 1$, the MI - SF
transition is of the ``ordinary'' type, with transverse (bath) teeth
ordering into LR SF and necessarily inducing LR SF order on the
(system) backbone. In contrast, for $J_\perp/J \ll 1$, the transition
is ``extra-ordinary'', with the backbone (a $d-d_\perp$ boundary for
the array of the transverse chains) ordering first, with $\psi \neq 0$
via a MI - SF$_{b}$ transition as $a$ drops below $c^2/a_\perp$, and
$\Psi(y)$ decays from $\Psi(y=0) = \psi_0\neq 0$ down to zero, with a
correlation length $\xi_\perp = \sqrt{K_\perp/a_\perp}$
\cite{MaxMetlitskiPRL22,
  RevisitedMaxMetlitski21}. This then is followed by another
SF$_{b}$-SF transition with $\Psi\neq 0$ as $a_\perp$ crosses
zero. This transition is MI - SF of an extra-ordinary type, where a
boundary at $y=0$ is already fully SF-ordered and remains so, as the
transverse bulk (teeth) order into a SF.

{\em Two-dimensional Luttinger model: phases and phase
  transitions.}---We now return to the 2+1D model of our interest,
Fig.\ref{latticeFig}, where, because of low-dimensionality and sparse
connectivity only quasi-long-range superfluid order is possible for
the decoupled transverse teeth (bath), requiring beyond-MFT treatment
above.  We now discuss the phases and the nature of quantum phase
transitions between them, illustrated in Fig.\ref{phasediagramFig}.


(i) For $\hat J_\alpha \ll 1$ and commensurate fillings
$\nu_\alpha = q_\alpha/p_\alpha$ ($q_\alpha,p_\alpha\in {\cal Z}$),
e.g., $\nu_\alpha = n_{0,\alpha}(\hat\mu)$ an integer
($\nu_\alpha=(\nu,\nu_\perp)$ labels backbone and transverse chain
fillings, respectively), the system is in a fully gapped MI state, as
in a conventional BH model.

(ii) For $J\gg J_\perp$, $\hat J$ above and $\Jh_\perp$ below their
lower-critical values, $\hat J_{c1}(\hat J_\perp,\hat \mu)$ and
$\hat J_{\perp,c1}(\hat J_,\hat \mu)$, (in mean-field theory
corresponding to $a < c^2/a_\perp$ and $a_\perp$ positive,
Eq.\ref{a_alpha}), respectively, at $T=0$ the MI undergoes a
Kosterlitz-Thouless (KT) MI-SF$_{b}$ transition to a quasi-long-ranged
superfluid along the $y=0$ $x$-backbone and short-range correlated
Mott-insulating transverse teeth-chains.  This is a regime of boundary
critical phenomena of extra-ordinary type~\cite{MaxMetlitskiPRL22,
  RevisitedMaxMetlitski21}, where the subsequent ordering of the bulk
(transverse comb chains) takes place in the presence of a QLRO
boundary (backbone).

Because the bulk (transverse teeth chains) remain gapped, this 1d
backbone boundary MI-LL$_{b}$ transition is well-captured by a
lattice-pinned Luttinger liquid, with a familiar Euclidean Lagrangian,
\bea
L_b = \int_x\left[\frac{1}{2\pi g} (\pt_\mu\theta_0)^2 - u\cos(2p\theta_0)\right],
\label{LagrLLb}
\eea
with $\theta_0(\tau,x)\equiv\theta(\tau,x,y=0)$ the backbone phonon
phase, the long wavelength density fluctuation,
$\delta n = -\frac{1}{\pi}\partial_x\theta_0$, and characterized by
the backbone Luttinger parameter $g(\Jh,\Jh_\perp,\hat\mu)$.  The
sine-Gordon model is well-known to undergo a `roughening''
(depinning) transition~\cite{GiamarchiBosonizationBook}, here
corresponding to the MI-LL$_b$ at $g_{c1}=2/p^2$, above which the
lattice potential $u$ becomes irrelevant.

(iii) Increasing $\hat J$ larger than
$\hat J_{c2}(\hat J_\perp,\hat \mu)$, SF$_{b}$ undergoes a quantum
transition to iSF, characterized by long-range SF order along $x$ at
$y=0$ backbone and quasi-long range SF order along $y$ transverse
teeth-chains. Although this unusual iSF phase was extensively studied
in Refs.~\onlinecite{iTQFprb, fingerprintsPRA} and is well understood,
the nature of the LL$_b$ - iSF ``extra-ordinary'' critical point
remains to be elucidated. The challenge is to include the interplay of
the backbone gapless superfluid fluctuations and the bulk KT QLR
critical modes of the transverse chains.

As analyzed in
Ref.~\onlinecite{MaxMetlitskiPRL22,RevisitedMaxMetlitski21}, we expect
two universality classes of this boundary LL$_b$ - iSF phase
transition, corresponding to irrelevant and relevant hybridization
coupling $c$ (in $S$ of \rfs{Sfull}), respectively, for
$\hat J < \hat J_c$-- ``extraordinary - LL'' -- and for
$\hat J > \hat J_c$ -- ``extraordinary - log'' sides of the special
multicritical point (red online) on the LL$_b$ - iSF phase boundary in
Fig.\ref{phasediagramFig}. To this end, simple analysis shows that the
RG eigenvalue (defined by $c(b) \sim c b^\lambda$ after
coarse-graining by a scale factor $b$) of the hybridization coupling
$c$ is $\lambda = 1-1/(4g) - 1/(4g_\perp)$. At the LL$_b$ - iSF phase
transition, $g_{\perp,c2} = 2/p_\perp^2$, $\lambda = 0$ gives
\bea
g_c &=& \frac{2}{8-p_\perp^2},
\label{gc}
\eea
as the multi-critical point (red dot in Fig. \ref{phasediagramFig}) on
the phase boundary separating two types of LL$_b$ - iSF phase
transition, where for $g < g_c$ ($g > g_c$) the backbone boundary is
decoupled (coupled) from the bulk transverse chains.  We then require
that $g_{c1} = 2/p^2 < g_c =2/(8-p_\perp^2) $, giving, for e.g.,
$p =1$, $p_\perp \ge 3$.

(iv) Conversely, for $J_\perp\gg J$, (in MFT
$a_\perp$ negative), for $\hat J_\perp > \hat J_{\perp c1}(\hat J,\hat
\mu)$ (corresponding to the Luttinger parameter $g_{\perp c1} =
2/p_\perp^2$), bosons undergo the MI -
LL$_\perp$ transition in a KT universality class, controlled by
\bea L_\perp = \sum_x\int_y\left[\frac{1}{2\pi
    g_\perp}(\pt_\mu\theta_x)^2 - u_\perp\cos(2p_\perp\theta_x)\right],
\label{LagrLLperp}
\eea
to a superfluid state that is quasi-long-range correlated along
$y$ of independent transverse LL chains, labelled by
$x$. We observe that, in contrast to higher dimensions (where
long-range transverse [bulk] chain SF order necessarily induces
long-range boundary order along $x$ in the
$y=0$ boundary backbone), in 2+1D a {\em quasi}-long range transverse
(bulk) order at sufficiently large
$p_\perp$ (low commensurability) allows for two distinct
possibilities: (a) short-range or (b) long-range correlated order
along $x$ on the
$y=0$ backbone, depending on the transverse chains filling
commensurability.  Thus, as illustrated in Fig. \ref{phasediagramFig},
we expect for $\hat J_{\perp c1}(\hat J,\hat \mu) < \hat J_\perp <
\hat J_{\perp c2}(\hat J,\hat \mu)$ and for $\hat J_{\perp c2}(\hat
J,\hat \mu) < \hat
J_\perp$ to have short-ranged and long-range SF orders along
$x$ at
$y=0$ backbone. The latter incoherent SF is the aforementioned iTQF
phase, previously extensively studied~\cite{iTQFprb, fingerprintsPRA}.
Both MI -- LL$_\perp$ at $\hat J_{\perp c1}(\hat J,\hat
\mu)$ and LL$_\perp$-- iSF at $\hat J_{\perp c2}(\hat J,\hat
\mu)$ are of the KT type.

To assess the possibility of this two-stage transition scenario (as in
main Fig.~\ref{phasediagramFig}), we analyze the relevance of the
backbone interchain hopping operator
$-J\sum_x\int_\tau\cos[\phi^0_{x+1}(\tau) - \phi^0_x(\tau)]$,
controlled by the Luttinger parameter
$g_\perp(\hat J_\perp,\hat J, \hat\mu)$ and $p_\perp$ of the
independent transverse LL chains. Standard analysis gives the RG
eigenvalue $\lambda_J = 1-1/(2g_\perp)$, that must be negative in the
range of $J_{\perp, c1}< J_{\perp} < J_{\perp c2}$ (corresponding to
$g_{\perp, c1}< g_{\perp} < g_{\perp c2}$) above the MI-LL$_\perp$
transition in order to have an independent transition at
$J_{\perp c2}$. This then requires that
$g_{\perp c2} = 1/2 > g_{\perp c1} = 2/p_\perp^2$, namely
$p_\perp > p_{\perp c} = 2$.  For higher level of commensurability,
$p_\perp \leq 2$, there is no distinct LL$_\perp$ phase and the phase
diagram is given by the inset of Fig.~\ref{phasediagramFig} with a
direct MI - iSF transition.

{\em Conclusions.}---We studied interacting bosons on a comb lattice
(Fig.~\ref{latticeFig}), mapping out the phase diagram in
Fig.~\ref{phasediagramFig}, detailing the nature of the phases --
Mott-insulator, backbone- and transverse-Luttinger liquids, and the
incoherent superfluid -- as well as the associated quantum phase
transitions, controlled by respective hopping amplitudes, on-site
interaction and the chemical potential (or equivalently filling
fractions).  Our predictions in this model should be testable by QMC
simulations\cite{ChaoZhangQMCunpublished}, and through experiments on
degenerate bosonic atoms in an optical lattice and holographic tweezer
arrays configured into a comb lattice (that may present some
experimental challenges~\cite{BakrQuench}). Our predictions should also be testable via
quantum simulation on an array of transmon qubits with Google's Willow
chip~\cite{PedramRoushan}, where Bose-Hubbard model emerges from the
nonlinear coupled Josephson-junction oscillators. As always,
sufficient cooling, coherence and lifetime will be the limiting
challenging factors to access our predictions.

Our results stimulate a number of interesting questions. These include
(i) post-quench dynamics across our phase diagram and for a variety of
initial conditions probing interacting quantum dynamics on the comb
lattice, (ii) inter-backbone-teeth entanglement entropy, and (iii)
boson transport along and across the backbone, and in particular how
it is modified by the transverse teeth and the associated phase
transitions.  We leave these many fascinating questions to a future
study.

{\it Acknowledgments}. L.R. thanks Anatoly Kuklov, Lode Pollet,
Nikolay Prokof’ev and Boris Svistunov for inspiring his interest in
this problem and earlier collaboration in Refs.~\onlinecite{iTQFprb,
  fingerprintsPRA}, and Adam Kaufman and Pedram Roushan for
discussions on experimental realizations.  L.R. also thanks Kavli
Institute for Theoretical Physics for its hospitality during a
sabbatical stay in Fall 2023 when this work was initiated. This work
was supported by the Simons Investigator Award to L.R. from the James
Simons Foundation and in part by the National Science Foundation under
Grant No. NSF PHY-1748958 and PHY-2309135.


\begin{thebibliography}{99}
\bibitem{commentFilling} The backbone $\nu = q/p$ and transverse-chain
  $\nu_\perp = q_\perp/p_\perp$ fillings (labelled by
  $\alpha = b, \alpha$) cannot be independently tuned with a single
  chemical potential $\mu$. However, a range of $\mu$, can give
  distinct but commensurate fillings $\nu$ and
  $\nu_\perp$. Furthermore, one can utilize single particle backbone
  and transverse-chain potentials to independently tune corresponding
  fillings.
\bibitem{commentOnSiteU} In our field theoretic analysis of the
  generic phase diagram, we generalize BH model interaction beyond
  on-site to allow for Mott-insulating phases pinned at low
  commensurabilty fillings.
\end{thebibliography}
\end{document}